\documentclass[a4paper]{article}

\usepackage{amsthm}
\usepackage{amsmath}
\usepackage{amssymb}
\usepackage{graphicx}
\usepackage{algorithm2e}
\usepackage{verbatim}
\usepackage{color}
\usepackage{tikz}

\usetikzlibrary{chains,matrix,scopes,decorations,%
shapes,arrows,shapes,positioning}
\tikzstyle{vertex}=[circle, draw, 
inner sep=0pt, minimum width=2pt]

\theoremstyle{plain}
\newtheorem{theorem}{Theorem}

\newtheorem{corollary}{Corollary}
\newtheorem{lemma}{Lemma}

\theoremstyle{definition}
\newtheorem{defi}{Definition}

\newenvironment{definition}[1][\noindent]
{\medskip \defi{\em (#1)}\\}
{\medskip \\\noindent}

\newenvironment{pf}
{\medskip\emph{Proof. }}
{\hfill$\Box$\medskip\noindent}

\newenvironment{pfof}[1]
{\medskip\emph{Proof (of {#1}). }}
{\hfill$\Box$\medskip\noindent}

\newcommand{\TUT}[3] {\mbox{$T({#1};{#2},{#3})$}}

\newcommand{\ZTUT}[3] {\mbox{$Z({#1};{#2},{#3})$}}

\newcommand{\ZNUL}[3] {\mbox{$\hat{Z}({#1};{#2},{#3})$}}

\newcommand{\REL}[2] {\mbox{$R({#1};{#2})$}}
\newcommand{\rel} {\mbox{$R$}}

\newcommand{\ZREL}[2]{\mbox{$\hat{R}({#1};{#2})$}}
\newcommand{\zrel} {\mbox{$\hat{R}$}}

\newcommand{\peval}[2]{{\sc \mbox{${#1}$}-val\mbox{$({#2})$}}}
\newcommand{\oeth}{{\footnotesize}{\sc eth}}
\newcommand{\ceth}{{\footnotesize\#}{\sc eth}}
\newcommand{\cp}{{\footnotesize\#}{\sc P}}

\newcommand{\inflate}[2] {\mbox{${#1}\otimes{#2}$}}

\newcommand{\bounce}[1] {\mbox{$B_{#1}$}}

\newcommand{\Oh}{O}

\begin{document}

\title{The Exponential Time Complexity of Computing the Probability That
a Graph is Connected\footnote{To appear in
\emph{5th International Symposium on Parameterized and Exact
  Computation (IPEC 2010)}, December 13–-15, 2010, Chennai, India,
Springer LNCS, 2010. Partially supported by Swedish Research Council grant VR 2007--6595.}}

\author{Thore Husfeldt
\footnote{IT University of Copenhagen, Denmark and Lund University,
Sweden} \and Nina Taslaman
\footnote{IT University of Copenhagen, Denmark}}
\date{}

\maketitle

\begin{abstract}
  We show that for every probability $p$ with $0<p<1$, computation of
  all-terminal graph reliability with edge failure probability $p$
  requires time exponential in $\Omega(m/\log^2 m)$ for simple graphs
  of $m$ edges under the Exponential Time Hypothesis.
\end{abstract}

\section{Introduction}

Graph reliability is a simple mathematical model of connectedness in
networks that are subject to random failure of its communication
channels. This type of stochastic networks arise naturally in, e.g.,
communication or traffic control; see \cite{BaCoPr95} for an extensive
survey of application areas.

For a connected graph $G=(V,E)$ and probability $p$, the {\em
  all-terminal reliability}  \REL{G}{p} is the probability that
there is a path of operational edges between every pair of nodes,
given that every edge of the graph fails independently with
probability $p$.
For example, with 
$p=\frac{1}{2}$, the all-terminal reliability of the graph
\newcommand{\fivecycle}{\begin{tikzpicture} \path (90+1*72:0.15) node[vertex]
      (v0) {}; \path (90+2*72:0.15) node[vertex] (v1) {}; \path
      (90+3*72:0.15) node[vertex] (v2) {}; \path (90+4*72:0.15)
      node[vertex] (v3) {}; \path (90+5*72:0.15) node[vertex] (v4) {};
      \draw (v0) -- (v1) -- (v2) -- (v3) -- (v4) -- (v0);
\end{tikzpicture}
}
$\vcenter{\hbox{$\fivecycle$}}$
is $\frac{3}{16}$, and the all-terminal reliability of the graph
$\vcenter{\hbox{$\begin{tikzpicture}
      \path (0:0) node[vertex] (c) {};
\path (45+0*90:0.2) node[vertex] (v1) {};
\path (45+1*90:0.2) node[vertex] (v2) {};
\path (45+2*90:0.2) node[vertex] (v3) {};
\path (45+3*90:0.2) node[vertex] (v4) {};
\draw (v2) -- (c) -- (v3) -- (v2);
\draw (v1) -- (c) -- (v4);
\end{tikzpicture}$}}$ is  $\frac{2}{16}$.

In general, for a connected, undirected graph $G=(V,E)$ with edge-failure
probability $p$, the all-terminal reliability can be given as
\begin{equation} \label{eq: reliability}
\REL{G}{p} =
\sum_{\substack{ A \subseteq E \\ spanning \\ connected }}
p^{|E\setminus A|}(1-p)^{|A|}
\,.
\end{equation}
For example $R( \vcenter{\hbox{$\fivecycle$}}\, ;\frac{1}{3}) =
(\frac{2}{3})^5 + 5\cdot \frac{1}{3}\cdot (\frac{2}{3})^4 =
\frac{112}{243}$. Computing \REL{G}{p} directly from \eqref{eq:
  reliability} can take up to $\Oh(2^m)$ operations, where $m$ is the
number of edges in $G$. An algorithm of Buzacott \cite{Buz80} solves
the problem in time $3^n n^{\Oh(1)}$ for graphs of $n$
vertices. Further improvements exist
\cite{BjHuKaKo08}, but remain exponential in $n$; this is explained 
by the $\exp(\Omega(n))$ lower bound of \cite{DeHuWa10}. On the other hand,
subexponential time algorithms have been found for some restricted
classes of graphs.  For example, the problem can be solved in time
$\exp( \Oh(\sqrt n))$ for planar graphs \cite{SeImTa95}. A natural
question is then whether the complexity can be reduced for for further 
classes of graphs. Especially, from an applications point of view, 
the case of simple graphs is interesting.

It is clear that $\REL{G}{0}=1$ and $\REL{G}{1}=0$ for all connected graphs, so
for some values of $p$ the problem is trivial. One may ask what the
situation is for values close to these extremes. Moreover, for
$p=\frac{1}{2}$ it is easily seen that \eqref{eq: reliability}
equals the number of connected, spanning subgraphs of $G$, divided by
$2^m$. This is an interesting enumeration problem in itself, and one
could be tempted to hope for a better algorithm than $\exp(O(n))$,
because a related enumeration problem, the number of spanning
\emph{trees} of $G$, can be solved in polynomial time by Kirchhoff's
matrix--tree theorem \cite{K}.

\subsection*{Result}

We give a lower bound on the problem of computing all-terminal graph
reliability for the class of simple graphs for all nontrivial
$p$, in the framework recently proposed by Dell \emph{et al.}
\cite{DeHuWa10}.  In particular, we work under the counting
exponential time hypothesis:
\begin{description}
\item[(\ceth)] There is a constant $c > 0$ such that no deterministic
algorithm can compute \#3-Sat in time $\exp(cn)$.
\end{description}
This is a relaxation of the exponential time hypothesis (\oeth) of
Impagliazzo \emph{et al.} \cite{IPZ}, so our results hold under
\oeth{} as well. The best current bound for \#3-Sat is $O(1.6423^n)$
\cite{Kut07}.

\begin{theorem}
  For any fixed probability $p$ with $0<p<1$, computing the all-terminal
  reliability \REL{G}{p} of a given simple graph $G$ of $m$ edges requires 
  time exponential in $\Omega(m/\log^2m)$ under \ceth.
\label{thm: 1}
\end{theorem}
In particular, the bound holds for $p=\frac{1}{2}$, i.e., counting the
number of connected spanning subgraphs of a given graph.

We have expressed the lower bound in terms of the parameter $m$, the number
of edges of the input graph. Since $n\leq m$ for connected graphs, the result
implies the lower bound $\exp(\Omega(n/\log^2n))$ in terms of the parameter
$n$, the number of vertices of the input graph. Moreover, the
$\Omega(m/\log^2m)$ lower bound together with the $\exp(O(n))$ algorithm from
\cite{Buz80,BjHuKaKo08} shows that the hard instances have roughly linear
density, ruling out a better algorithm than $\exp(O(n/\log^2 n))$ also for
the restricted case of sparse graphs.

Our bound does not quite match the best known upper bound $\exp(O(n))$
of \cite{Buz80,BjHuKaKo08}. This situation is similar to the bounds reported
in \cite{DeHuWa10} for related problems on simple graphs, which also
fall a few logarithmic factors (in the exponent) short of the best
known algorithms. The bound does, however, suffice to separate the
complexity of reliability computation from the $\exp( \Oh(\sqrt n))$
bound for the planar case \cite{SeImTa95}.

\subsection*{Graph polynomials}
Expression~\eqref{eq: reliability}, viewed as a function $p\mapsto \REL{G}{p}$
for fixed $G$, is known as the {\em reliability polynomial} of $G$, an
object studied in algebraic graph theory \cite[Sec.~15.8]{GoRo01}. For
example, $R(\vcenter{\hbox{\fivecycle}}\!;p) = (1-p)^5 + 5p(1-p)^4$.

Arguably, the most important graph polynomial is the bivariate {\em
  Tutte polynomial} \TUT{G}{x}{y}, which encodes numerous
combinatorial parameters of the input graph $G$, and whose restriction
to certain lines and curves in the $xy$-plane specialize to other well
known graph polynomials. The reliability polynomial is essentially a
restriction of this polynomial to the ray $\{\,(1,y) \colon
y> 1\,\}$. The complexity of computing \TUT{G}{x}{y} at various points
$(x,y)$ with respect to an input graph $G$ is very well-studied in
various models of computation, and the present paper thus establishes
lower bounds for simple graphs along the mentioned ray, which was left
open in a recent study \cite{DeHuWa10} to completely map the
exponential time complexity of the ``Tutte plane''.

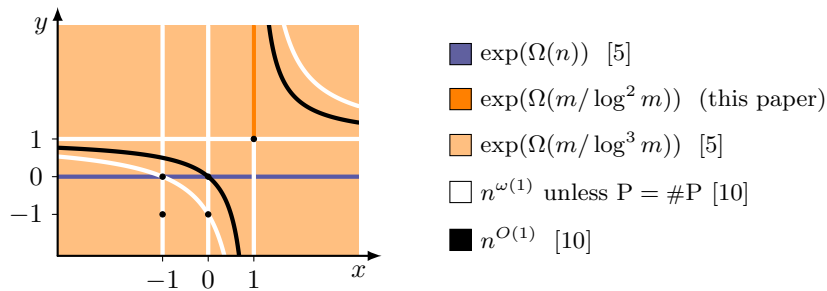
\begin{figure}
\[
\vcenter{\hbox{
\begin{tikzpicture} [line cap=round,line join=round,x=.6cm,y=.5cm]
  \def\xmin{-3.3}; \def\ymin{-2.1};\def\xmax{3.3}; \def\ymax{4.0};

  \begin{scope}
  \clip (\xmin,\ymin) rectangle (\xmax,\ymax);

  \draw[fill, orange!50]  (\xmin,\ymin) rectangle (\xmax, \ymax);
   \draw[color=blue!30!gray, ultra thick] (\xmin,0)--(\xmax,0);
   \draw[color=white, ultra thick] (\xmin,1)--(\xmax,1);
   \draw[color=white, ultra thick] (1,\ymin)--(1,\ymax);
   \draw[color=white, ultra thick] (0,\ymin)--(0,\ymax);
   \draw[color=white, ultra thick] (-1,\ymin)--(-1,\ymax);

   \draw[color=orange, ultra thick] (1,1)--(1,\ymax);
   
  \draw[smooth,ultra thick,samples=100,domain=\xmin:.9] plot(\x,{1+1/(\x-1)});
  \draw[smooth,ultra thick,samples=100,domain=1.1:\xmax] plot(\x,{1+1/(\x-1)});

  \draw[smooth,ultra thick,white,samples=100,domain=\xmin:.9] plot(\x,{1+2/(\x-1)});
  \draw[smooth,ultra thick,white,samples=100,domain=1.1:\xmax] plot(\x,{1+2/(\x-1)});

   \draw[fill] (-1,-1) circle (1pt);
   \draw[fill] (-1,0) circle (1pt);
   \draw[fill] (0,-1) circle (1pt);
   \draw[fill] (0,0) circle (1pt);
   \draw[fill] (1,1) circle (1pt);
  \end{scope}
  \draw[thick,-latex] (\xmin,\ymin) -- (\xmax+.5,\ymin);
  \draw[shift={(\xmax,\ymin)}] node[below] {$x$};
  \foreach \x in {-1,0,1}
  \draw[shift={(\x,\ymin)}] (0pt,0pt) -- (0pt,-2pt)
                                   node[below] {$\x$};

  \draw[thick,-latex] (\xmin,\ymin) -- (\xmin,\ymax+.5);
  \draw[shift={(\xmin,\ymax)}] node[left] {$y$};
  \foreach \y in {-1,0,1}
      \draw[shift={(\xmin,\y)}] (0pt,0pt) -- (-2pt,0pt)
                                    node[left] {$\y$};
\end{tikzpicture}}}
\qquad
\newcommand\legend[2]{
  \tikz\node[fill=#1,draw,inner sep=4,label=right:{#2}] {}; 
  }
\vcenter{\hbox{\small
$\begin{array}{l}
  \legend{blue!25!gray}{$\exp(\Omega( n))$ 
		\text{  \cite{DeHuWa10}}}		\\[-2pt]
  \legend{orange}{$\exp(\Omega(m/\log^2m))$ 
		\text{  (this paper)} }			\\[-2pt]
  \legend{orange!50}{$\exp(\Omega(m/\log^3 m))$ 
		\text{  \cite{DeHuWa10}} }		\\[-2pt]
  \legend{white}{$n^{\omega(1)}$
		\text{unless $\text{P}= \text{\cp{}}$ \cite{JaVeWe90}}}	\\[-2pt]
  \legend{black}{$n^{O(1)}$ 
		\text{  \cite{JaVeWe90}}}
\end{array}$}}
\]
\caption{\label{fig: Tutte}\small Exponential time complexity under
  \ceth{} of the Tutte plane for simple graphs.}
\end{figure}

\subsection*{Related work}

The structural complexity of all-terminal graph reliability was
studied by Provan and Ball \cite{PrBa83}. For any probability $p$,
with $0<p<1$,
it is shown that computing $\REL{G}{p}$ for given $G$ is hard for
Valiant's counting class \cp{} \cite{Val79}.  The reductions in
\cite{PrBa83} do not preserve the parameters $n$ and $m$, so that the
running time bounds under \ceth{} implicitly provided by their
techniques are typically exponential in $\Omega(n^{1/k})$ for some
$k$.

The reliability problem under consideration in this paper 
admits a number of natural extensions.
\begin{enumerate}
\item We can consider the computational problem of finding the
  reliability polynomial itself, instead of its value at a fixed point
  $p$.  The input to this problem is a graph, and the output is a list
  of coefficients. For example, on input $\vcenter{\hbox{\fivecycle}}$  
  the output should give $R(\vcenter{\hbox{\fivecycle}}\!;p) = 
  4 p^5-15 p^4+20 p^3-10 p^2+1$.
\item We can associate individual probabilities to every
edge. For example, the graph $\vcenter{\hbox{\begin{tikzpicture} \draw
      node[vertex] (u) at (0,0) {}; \draw node[vertex] (v) at (.4,0)
      {}; \draw node[vertex] (w) at (.8,0) {}; \draw (u)-- node[above]
      {$\scriptstyle\frac{1}{2}$} (v) -- node[above] {$\scriptstyle
        \frac{1}{4}$} (w);
\end{tikzpicture}}}$
becomes disconnected with probability $\frac{5}{8}$.
\item We can consider multigraphs like
\begin{tikzpicture}
\draw node[vertex] (u) at (0,0) {};
\draw node[vertex] (v) at (.4,0) {};
\draw node[vertex] (w) at (.8,0) {};
\draw (u) -- (v);
\draw (v) to [bend left] (w);
\draw (v) to [bend right] (w);
\end{tikzpicture}, but with the same edge weight $p$. As indicated by
the examples (for $p=\frac{1}{2}$), the multigraph case is a special
case of the individually edge-weighted case, a fact that we will use
later.
\end{enumerate}
All of these problems are at least as hard as the problem under
consideration in the present paper. Lower bounds of size
$\exp(\Omega(m))$ are given in \cite{DeHuWa10} or follow relatively
easily (see \S\ref{sec: coefficients}).

\smallskip A recent paper of Hoffman \cite{CH10} studies the complexity
of another graph polynomial, the independent set polynomial, in the
same framework.

\section{Preliminaries}

We will only be concerned with undirected graphs. For a  graph
$G=(V,E)$ let $n$ denote the number of vertices, $m$ the number of edges,
and for any subset $A\subseteq E$ let $\kappa(A)$ denote the number of
connected components in the subgraph $(V,A)$ (especially, $\kappa(A)=1$
means that the edge subset $A$ is spanning and connected). Also, for graph
polynomials $P$ and $Q$, we write $P(G;\mathbf x) \sim Q(G';\mathbf x')$ if
the two expressions are equal up to an easily computable factor.

\subsection{Weighted reliability}

The reliability polynomial can be formulated as a restriction of the
{\em Tutte polynomial}, which for an undirected graph $G$ is given by
\[
\TUT{G}{x}{y} =
\sum_{A \subseteq E} (x-1)^{\kappa(A)-\kappa(E)}
(y-1)^{\kappa(A)+|A|-|V|}
\,.
\]
Note that $\kappa(E)=1$ in our case. We find the reliability polynomial
along the ray $\{\,(1, y)\colon y >1\,\}$ in the so called {\em Tutte plane}, as
$\REL{G}{p} \sim \TUT{G}{1}{1/p}$; in full detail:
\begin{equation}\label{eq: rel-to-tut}
	\REL{G}{p} = p^{m-n+1}(1-p)^{n-1} \TUT{G}{1}{1/p}\,,\qquad (0<p<1)\,.
\end{equation}

For complexity analysis of the Tutte polynomial, it has proved a
considerable technical simplification to consider Sokal's {\em
multivariate Tutte polynomial} \cite{Sok05}. Here the graph is equipped
with some weight function $\mathbf w \colon E \rightarrow \mathbb{R}$,
and the polynomial is given by
\[
\ZTUT{G}{q}{\mathbf w} =
\sum_{A\subseteq E} \mathbf w(A) q^{\kappa(A)}
\,,
\]
where $\mathbf w(A)=\prod_{e\in A}\mathbf w(e)$ is the edge-weight product
of the subset $A$. For constant edge weights $w = y-1$ we have
$\ZTUT{G}{q}{w} \sim \TUT{G}{x}{y}$ with $q=(x-1)(y-1)$. The
``reliability line'' $x=1$ in the Tutte plane thus corresponds to $q=0$
in the weighted setting, where $Z$ vanishes, so instead we will consider
the slightly modified polynomial
\medskip
\[
	\ZNUL{G}{q}{\mathbf w} =
	q^{-1} \ZTUT{G}{q}{\mathbf w}
	\,.
\]

\medskip\noindent
At $q=0$, this gives a weighted version of the reliability polynomial:
\begin{definition}[Weighted reliability polynomial]
For a connected, undirected graph $G=(V,E)$, the
{\em weighted reliability polynomial} of $G$ is given by
\begin{equation}\label{eq: r-multi}
\ZREL{G}{\mathbf w}
\; =\;
\ZNUL{G}{q}{\mathbf w}|_{q=0}
\;=
\sum_{\substack{ A \subseteq E \\ \kappa(A)=1 }}
\mathbf w(A)
\,.
\end{equation}
\end{definition}
For constant edge weight $w>0$ we have $\ZREL{G}{w} = w^{n-1}
\TUT{G}{1}{1+w}$, so for
  $0<p<1$ we can recover the reliability polynomial
through \eqref{eq: rel-to-tut} as
\begin{equation} \label{eq: rel-to-zrel}
	\REL{G}{p} = p^m \ZREL{G}{1/p-1}\,.
\end{equation}

\subsection{Graph transformations}

A classical technique for investigating the complexity of the Tutte
polynomial at a certain point $(x',y')$ of the Tutte plane, is to relate
it to some already settled point $(x,y)$ via a graph transformation
$\varphi$, such that $\TUT{G}{x'}{y'} \sim \TUT{\varphi(G)}{x}{y}$. For
the weighted setting we have the following rules, which are simple
generalizations of \cite[Sec. 4.3]{GoJe08}. (See
Appendix~\ref{ap: pf_thickstretch}.) For a graph $G=(V,E)$ with edge
weights given by $\mathbf w$:
\begin{lemma}
If $\varphi(G)$ is obtained from $G$ by replacing a single edge
$e\in E$ with a simple path of $k$ edges $P=\{e_1,...,e_k\}$
with $\mathbf w(e_i)= w_i$, then
\[
\ZREL{\varphi(G)}{\mathbf w} =
C_P \cdot \ZREL{G}{\mathbf w[e\mapsto w']}
\,,
\]
where
\[
\frac{1}{w'} =
\frac{1}{w_1}+\cdots+\frac{1}{w_k}
\;\;\;\;\; \text{ and } \;\;\;\;\;
C_P = \frac{1}{w'}\prod_{i=1}^k w_i
\,.
\]
\label{lem: stretch}
\end{lemma}
\begin{lemma}
If $\varphi(G)$ is obtained from $G$ by replacing a single edge
$e\in E$ with a bundle of parallel edges $B
=\{e_1,\ldots,e_k\}$ with $\mathbf w(e_i)= w_i$, then
  \[
\ZREL{\varphi(G)}{\mathbf w} =
\ZREL{G}{\mathbf w[e\mapsto w']}
\,,
\]
where
  \[
w' = -1 + \prod_{i=1}^k (1+w_i)
\,.
\]
\label{lem: thick}
\end{lemma}
\begin{corollary} \label{cor: stretchthick}
If $\varphi(G)$ is obtained from $G$ by replacing a single edge
$e\in E$ with a simple path of $k$ edges of constant weight $w$,
then
\begin{equation} \label{eq: stretch}
\ZREL{\varphi(G)}{\mathbf w} =
kw^{k-1} \cdot \ZREL{G}{\mathbf w[e\mapsto w/k]}
\,,
\end{equation}
and if it is obtained from $G$ by replacing $e\in E$ with a
bundle of $k$ parallel edges of constant weight $w$, then
\begin{equation} \label{eq: thick}
\ZREL{\varphi(G)}{\mathbf w} =
\ZREL{G}{\mathbf w[e\mapsto (1+w)^k-1]}
\,.
\end{equation}
\end{corollary}
These rules are transitive \cite[Lem. 1]{GoJe08}, and so can be freely
combined for more intricate weight shifts. To preserve constant weight
functions we need to perform the same transformation to every edge of the
graph. This calls for the graph theoretic version of Brylawski's tensor
product for matroids \cite{Bry80}. We found the following terminology
more intuitive for our setting:
\begin{definition}[Graph inflation]
  Let $H$ be a 2-terminal undirected graph. For any undirected graph
  $G=(V,E)$, an {\em $H$-inflation} of $G$, denoted \inflate{G}{H},
  is obtained by replacing every edge $xy\in E$ by (a fresh copy of)
  $H$, identifying $x$ with one of the terminals of $H$ and $y$ with
  the other.\footnote{
This can, in general, be done in two different ways, resulting
in graphs that need not be isomorphic. However, the Tutte
polynomial is blind to this difference.
See extensive footnote in \cite{DeHuWa10}, Section 5.1.
}
\end{definition}
If $H$ is a simple path of $k$ edges, \inflate{G}{H} gives the {\em
$k$-stretch} of $G$. Similarly, a bundle of $k$ parallel edges results in
a {\em $k$-thickening}, of $G$.

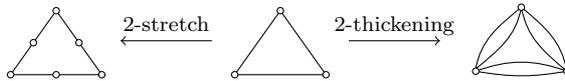
\begin{center}
\begin{tikzpicture}[scale=0.3]
\draw (0,0) -- (2,2.82842712) -- (4,0) -- (0,0)
plot[only marks, mark=*, mark options={fill=white}, mark size=4pt]
coordinates{
(0,0)(1,2.82842712/2)(2,2.82842712)(3,2.82842712/2)(4,0)(2,0)
};
\end{tikzpicture}
\begin{tikzpicture}[scale=0.3]
	\draw [->] (4,1.5) -- (0,1.5);
	\coordinate (label) at (2,1.5);
	\node [above] at (label) {\footnotesize 2-stretch};
	
	\draw (5+0,0) -- (5+2,2.82842712) -- (5+4,0) -- (5+0,0)
	plot[only marks, mark=*, mark options={fill=white}, mark size=4pt]
	coordinates{(5+0,0)(5+2,2.82842712)(5+4,0)};
	
	\draw [->] (10,1.5) -- (10+4,1.5);
	\coordinate (label) at (10+2,1.35);
	\node [above] at (label) {\footnotesize 2-thickening};
\end{tikzpicture}
\begin{tikzpicture}[scale=0.3]
	\draw (0,+0.1) to [out=80, in=220] (2,2.82842712)
	to [out=-40,in=100] (4,0)
	to [out=190,in=-10] (0,0);
	\draw (0,+0.1) to [out=40, in=260] (2,2.82842712)
	to [out=-80,in=140] (4,0)
	to [out=160,in=20] (0,0)
	plot[only marks, mark=*, mark options={fill=white}, mark size=4pt]
	coordinates{(0,0)(2,2.82842712)(4,0)};
\end{tikzpicture}
\end{center}

\subsection{Hardness of computing coefficients}
\label{sec: coefficients}

Our pivot for proving Theorem~\ref{thm: 1} will be the following hardness
result, which says that even when restricted to fixed hyperbolas
$(x-1)(y-1)=q$, computing the full Tutte polynomial is hard. This is an
extension to the case $q=0$ of Lemma~2 in \cite{DeHuWa10}, and the proof
is given in Appendix~2.
\begin{lemma}
   Under \ceth, computing the coefficients of the polynomial
$w\mapsto \ZNUL{G}{q}{w}$ for given simple graph $G$ and
rational number $q\notin \{1,2\}$ requires time
exponential in $\Omega(m)$.
\label{lem: coeff}
\end{lemma}
Since \REL{G}{p} is essentially \ZREL{G}{\mathbf w} restricted to
positive constant weight functions, and since $\ZREL{G}{\mathbf w} =
\ZNUL{G}{0}{\mathbf w}$, the following is immediate:
\begin{corollary} \label{corr: pivot}
Under \ceth, it requires time exponential in $\Omega(m)$ to compute
the coefficients of the reliability polynomial.
\end{corollary}

\section{Bounce graphs}

As a first step towards Theorem~\ref{thm: 1}, we present here a class of
graph transformations whose corresponding weight shifts for the
reliability polynomial are all distinct. These transformations are mildly
inspired by $k$-byte numbers, in the sense that each has associated to it
a sequence of length $k$, such that the lexicographic order of these
sequences determines the numerical order of the corresponding (shifted)
weights. Each transformation is a {\em bounce inflation}:
\begin{definition}[Bounce graph]
For positive numbers $h$ (height) and $l$ (length), the
{\em $(h,l)$-bounce} is the graph obtained by identifying all the
left and all the right endpoints of $h$ simple paths of length $l$.
Given a {\em bounce sequence},
$S = \langle s_1, s_2, \dots, s_k \rangle$,
of $k$ numbers $s_i>1$, the corresponding {\em bounce graph},
$\bounce{S}$, is the (simple) graph obtained by concatenating $k$
$(h,l)$-bounces by their endpoints, where the height starts at $1$
for the first bounce and then increases by one for each follower,
and the length of the $i$th bounce is $s_i$.

{\centering
\hfill
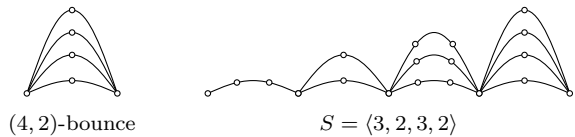
\par
\begin{tikzpicture}[scale=0.85]
\draw (0-1.4,0) sin (0.7-1.4,0.2) cos (1.4-1.4,0);
\draw (0-1.4,0) sin (0.7-1.4,0.6) cos (1.4-1.4,0);
\draw (0-1.4,0) sin (0.7-1.4,0.95) cos (1.4-1.4,0);
\draw (0-1.4,0) sin (0.7-1.4,1.3) cos (1.4-1.4,0)
plot[only marks, mark=*, mark options={fill=white}, mark size=1.2pt]
coordinates{ (0-1.4,0) (0.7-1.4,0.6) (0.7-1.4,0.2)
(0.7-1.4,0.95) (0.7-1.4,1.3) (1.4-1.4,0)};
\coordinate (label) at (0.7-1.4,-0.8);
\node [above] at (label) {\footnotesize $(4,2)$-bounce};

\draw (1*1.4,0) sin (1*1.4+0.7,0.2) cos (1*1.4+1.4,0)
plot[only marks, mark=*, mark options={fill=white}, mark size=1.2pt]
coordinates{(1*1.4+0,0) (1*1.4+0.45,0.17) (1*1.4+0.95,0.17) (1*1.4+1.4,0) };
\draw (2*1.4,0) sin (2*1.4+0.7,0.2) cos (2*1.4+1.4,0);
\draw (2*1.4,0) sin (2*1.4+0.7,0.6) cos (2*1.4+1.4,0)
plot[only marks, mark=*, mark options={fill=white}, mark size=1.2pt]
coordinates{(2*1.4+0,0) (2*1.4+0.7,0.2) (2*1.4+0.7,0.6) (2*1.4+1.4,0)};
\draw (3*1.4,0) sin (3*1.4+0.7,0.2) cos (3*1.4+1.4,0);
\draw (3*1.4,0) sin (3*1.4+0.7,0.6) cos (3*1.4+1.4,0);
\draw (3*1.4,0) sin (3*1.4+0.7,0.95) cos (3*1.4+1.4,0)
plot[only marks, mark=*, mark options={fill=white}, mark size=1.2pt]
coordinates{
(3*1.4,0) (3*1.4+1.4,0)
(3*1.4+0.45,0.17) (3*1.4+0.95,0.17)
(3*1.4+0.43,0.50) (3*1.4+0.97,0.50)
(3*1.4+0.41,0.77) (3*1.4+0.99,0.77)
};
\draw (4*1.4,0) sin (4*1.4+0.7,0.2) cos (4*1.4+1.4,0);
\draw (4*1.4,0) sin (4*1.4+0.7,0.6) cos (4*1.4+1.4,0);
\draw (4*1.4,0) sin (4*1.4+0.7,0.95) cos (4*1.4+1.4,0);
\draw (4*1.4,0) sin (4*1.4+0.7,1.3) cos (4*1.4+1.4,0)
plot[only marks, mark=*, mark options={fill=white}, mark size=1.2pt]
coordinates{
(4*1.4+0,0) (4*1.4+0.7,0.2) (4*1.4+0.7,0.6)
(4*1.4+0.7,0.95) (4*1.4+0.7,1.3) (4*1.4+1.4,0)
};
\coordinate (label) at (3*1.4,-0.8);
\node [above] at (label)
{\footnotesize $S=\langle 3,2,3,2 \rangle$};
\end{tikzpicture}
\par}
\noindent The {\em length} of a bounce graph is the number of bounces
in it (or, equivalently, the height of the highest bounce).
\end{definition}
Inflation by a bounce graph has the following weight-shifting effect,
from \zrel's perspective:
\begin{lemma}
For a graph $G$, bounce sequence
$S = \langle s_1, s_2, \dots, s_k \rangle$ and $w > 0$:
   \[
\ZREL{\inflate{G}{\bounce{S}}}{w} =
C_S^m \cdot \ZREL{G}{w_S}
\,,
\]
   where
\[
\frac{1}{w_S} =
\sum_{i=1}^{k} \frac{1}{(1+w/s_i)^i-1}
\;\;\;
\text{ and }
\;\;\;
C_S =
\frac{1}{w_S}
\cdot
\prod_{i=1}^k
w^{(s_i-1)i}
\left(
(w+s_i)^i-s_i^i
\right)
\,.
\]
\label{lem: bounceweights}
\end{lemma}
\begin{pf}
Starting out with \inflate{G}{\bounce{S}},
we will look at the effect of replacing one of the $m$ copies of
$\bounce{S}$ with a single edge $e$. We show that, with $\varphi$
denoting this operation:
\begin{equation} \label{eq: flatbounce}
\ZREL{\inflate{G}{\bounce{S}}}{w} =
C_S \cdot
\ZREL{\varphi (\inflate{G}{\bounce{S})}}
{\mathbf w[e\mapsto w_S]}
\,,
\end{equation}
where $w_S$ has the above form, and $\mathbf w$ has the old value $w$
on all unaffected edges. The lemma follows from performing
$\varphi$ for every copy of $\bounce{S}$ in \inflate{G}{\bounce{S}}.

The first step towards transforming a bounce graph (say,
\begin{tikzpicture}[scale=0.38]
\draw (0*1.4,0) sin (0*1.4+0.7,0.2) cos (0*1.4+1.4,0)
plot[only marks, mark=*, mark options={fill=white}, mark size=2.5pt]
coordinates{(0*1.4+0,0) (0*1.4+0.45,0.17) (0*1.4+0.95,0.17) (0*1.4+1.4,0) };
\draw (1*1.4,0) sin (1*1.4+0.7,0.2) cos (1*1.4+1.4,0);
\draw (1*1.4,0) sin (1*1.4+0.7,0.6) cos (1*1.4+1.4,0)
plot[only marks, mark=*, mark options={fill=white}, mark size=2.5pt]
coordinates{(1*1.4+0,0) (1*1.4+0.7,0.2) (1*1.4+0.7,0.6) (1*1.4+1.4,0)};
\draw (2*1.4,0) sin (2*1.4+0.7,0.2) cos (2*1.4+1.4,0);
\draw (2*1.4,0) sin (2*1.4+0.7,0.6) cos (2*1.4+1.4,0);
\draw (2*1.4,0) sin (2*1.4+0.7,0.95) cos (2*1.4+1.4,0)
plot[only marks, mark=*, mark options={fill=white}, mark size=2.5pt]
coordinates{
(2*1.4,0) (2*1.4+0.7,0.2) (2*1.4+0.7,0.6) (2*1.4+0.7,0.95)  (2*1.4+1.4,0)};
\end{tikzpicture})
into a single edge, is to replace each path of each bounce in it by a
single edge. Applying \eqref{eq: stretch} of Corollary~\ref{cor:
stretchthick} to each path of the $i$th bounce gives a factor
$(s_iw^{s_i-1})^i$ to the polynomial, and each edge in the resulting
$(h,1)$-bounce gets weight $w/s_i$ in the modified graph. Repeating
this process for every bounce gives a simplified bounce graph
(\begin{tikzpicture}[scale=0.35]
\draw (0*1.4,0) sin (0*1.4+0.7,0.2) cos (0*1.4+1.4,0)
plot[only marks, mark=*, mark options={fill=white}, mark size=2.5pt]
coordinates{(0*1.4+0,0) (0*1.4+1.4,0) };
\draw (1*1.4,0) sin (1*1.4+0.7,0.2) cos (1*1.4+1.4,0);
\draw (1*1.4,0) sin (1*1.4+0.7,0.6) cos (1*1.4+1.4,0)
plot[only marks, mark=*, mark options={fill=white}, mark size=2.5pt]
coordinates{(1*1.4+0,0) (1*1.4+1.4,0)};
\draw (2*1.4,0) sin (2*1.4+0.7,0.2) cos (2*1.4+1.4,0);
\draw (2*1.4,0) sin (2*1.4+0.7,0.6) cos (2*1.4+1.4,0);
\draw (2*1.4,0) sin (2*1.4+0.7,0.95) cos (2*1.4+1.4,0)
plot[only marks, mark=*, mark options={fill=white}, mark size=2.5pt]
coordinates{(2*1.4,0) (2*1.4+1.4,0)};
\end{tikzpicture})
in a transformed graph $\phi(\inflate{G}{\bounce{S}})$ such that
\[
\ZREL{\inflate{G}{\bounce{S}}}{w} =
\left(\prod_{i=1}^k (s_iw^{s_i-1})^i \right)
\cdot
\ZREL{\phi(\inflate{G}{\bounce{S}})}{\mathbf w'}
\,,
\]
with $\mathbf w'$ taking the value $w/s_i$ for every edge in the
$i$th bounce of the simplified bounce graph, and the old value $w$
outside it. Next, we successively replace each of its $(h,1)$-bounces
by a single edge, to get a simple path
(\begin{tikzpicture}[scale=0.35]
\draw (0,0) -- (1.4,0) -- (2*1.4,0) -- (3*1.4,0)
plot[only marks, mark=*, mark options={fill=white}, mark size=3pt]
coordinates{(0*1.4,0) (1*1.4,0) (2*1.4,0) (3*1.4,0)};
\end{tikzpicture})
of length $k$ (with non-constant edge weights).
From \eqref{eq: thick} of Corollary~\ref{cor: stretchthick}, we
know that this does not produce any new factors for the polynomial,
but the weight of the $i$th edge in this path will be given by
\[
w_i=
\left(
1+w/s_i
\right)^i - 1
\,.
\]
Finally, we compress the path into a single edge $e$. A single
application of Lemma~\ref{lem: stretch} then gives the result in
\eqref{eq: flatbounce}.
\end{pf}

If Lemma~\ref{lem: coeff} is our pivot for proving Theorem~\ref{thm: 1},
the following result is the lever:
\begin{lemma} \label{lem: lever}
For any size $m$, there exist $m+1$ distinct, simple bounce graphs
$\bounce{S}$ of size $\Oh(\log^2 m)$,
such that for any two associated bounce sequences $S$ and $T$:
\begin{equation} \label{eq: seqorder}
S >_{lex} T \; \Rightarrow \; w_{S} < w_{T}
\end{equation}
for all $w>6$.
\end{lemma}
\begin{pf}
The set of bounce sequences $S=\langle s_1,\dots,s_l\rangle$
of length $l=\log(m+1)$ and with each $s_i \in \{2,3\}$,
provides $m+1$ different simple bounce graphs of promised size.
To show that any two of them satisfy equation~\eqref{eq: seqorder}
we will look at the difference
\[
\Delta_{S,T}(w) = \frac{1}{w_S}-\frac{1}{w_T}
\,,
\]
and show that $\Delta_{S,T}(w)>0$ for $w>6$.

Let $k$ be the first index where the sequences differ,
say $s_k=3$ and $t_k=2$. We then have
\begin{eqnarray*}
\Delta_{S,T}(w)
& =
& \frac{1}{(1+w/3)^k-1} +
\sum_{i=k+1}^{l} \frac{1}{(1+w/s_i)^i-1} \nonumber \\
& -
& \frac{1}{(1+w/2)^k-1}
- \sum_{i=k+1}^{l} \frac{1}{(1+w/t_k)^i-1} \,.
\nonumber\\
\end{eqnarray*}
This would be minimal if $s_i=2$ and $t_i=3$ for all $i>k$, i.e. if
\[
\Delta_{S,T}(w) =
f(1+w/3) - f(1+w/2)
\,,
\]
where
\[
f(x) =
\frac{1}{x^k-1} -
\sum_{i=k+1}^{l} \frac{1}{x^i-1}
\,.
\]
If we could show that $f'(x)<0$ for $x>x_0$ then it would follow
(e.g. from the mean value theorem) that $f(x)>f(y)$ for $x<y$ above
$x_0$. In particular, with $x_0=3$ this would prove our claim. To
see that this is indeed the case, we look at the derivative
\[
f'(x) =
-\frac{kx^{k-1}}{(x^k-1)^2} +
\sum_{i=k+1}^{l} \frac{ix^{i-1}}{(x^i-1)^2}
\,.
\]
A bit of manipulation shows that the terms of the sum,
let us call them $T_i$, satisfy $T_i> 2 T_{i+1}$ for $x>3$, so
\[
f'(x)
<
\frac{kx^{k-1}}{(x^k-1)^2}
\left(
-1 + \sum_{i=k+1}^{l} \frac{1}{2^i}
\right)
<
0
\]
for $x>3$, and we are done by the above argument.
\end{pf}

\section{Evaluating the reliability polynomial is hard}

We are ready to prove Theorem~\ref{thm: 1}. We introduce
the following  notation for the problem of evaluating a graph
polynomial $P(G;\mathbf x)$ at a given point $\mathbf x$

\begin{quote} \peval{P}{\mathbf x}:

\begin{tabular}{rl}
{\bf input} & A simple, connected, undirected graph $G$ \\
{\bf output} & The value of $P(G;\mathbf x)$ (a rational number)
\end{tabular}
\end{quote}
In this notation, the computational problem described in Theorem~\ref{thm:
1} can be written as \peval{\rel}{p}, and the corresponding problem for
weighted reliability with constant edge weight $w$ is \peval{\zrel}{w}. We
will prove Theorem~\ref{thm: 1} by reducing the problem of computing
coefficients of the polynomial $p \mapsto \REL{G}{p}$, to the problem
\peval{\rel}{p} for any arbitrary fixed probability $p$ with $0<p<1$.

\begin{pfof}{Theorem~\ref{thm: 1}}
Let $G$ be a simple graph with $n$ vertices and $m$ edges. We prove that
\peval{\zrel}{w} requires time exponential in $\Omega(m/\log^2m)$ for any
$w>0$, which by \eqref{eq: rel-to-zrel} gives the same bound for
\peval{\rel}{p} for any $p$ with $0<p<1$.

Suppose we have an algorithm for \peval{\zrel}{w} for some fixed
$w>0$. For simplicity of exposition, first assume $w>6$. From
Lemma~\ref{lem: lever} we can easily construct $m+1$ bounce graphs
\bounce{S} such that each \ZREL{\inflate{G}{\bounce{S}}}{w}
gives us the value of \ZREL{G}{w_S} at some new weight $w_S$, with
all such $w_S$ distinct. Computing \peval{\zrel}{w} for each of these
$m+1$ bounce inflations, we get the value of $\ZREL{G}{w}$ at $m+1$
distinct $w$-values. Since the degree of this polynomial is $m$, that
gives us the coefficients by interpolation. By
Corollary~\ref{corr: pivot}, the whole process must then require time
$\exp(\Omega(m))$. By Lemma~\ref{lem: lever}, each inflation
\inflate{G}{\bounce{S}} will have $\Oh(m \log^2 m)$ edges.
Thus, for graphs of size $\Oh(m \log^2 m)$, the problem
\peval{\zrel}{w} has a lower bound of
$\exp(\Omega(m))$. The claimed bound then follows from the fact that
\[
\varphi(m) = m \log^2 m \;\; \implies \;\; \varphi^{-1}(m) \in
\Oh \left(m/\log^2 m\right)
\,.
\]

We turn to the case $0<w\leq 6$. Given such a $w$, choose a number $k$
such that
\[
w' := (w/2+1)^k-1 > 6 \,.
\]
From the above, we know that for any simple graph $G$ it takes time
exponential in $\Omega(m/\log^2 m)$ to evaluate \ZREL{G}{w'}. Now consider
the $k$-thickening-$2$-stretch of $G$, let us call this $G'$. This will be
a simple graph of $\Oh(km)$ edges, and from Corollary~\ref{cor:
stretchthick} it follows that the value of \ZREL{G'}{w} would give us the
value of \ZREL{G}{w'}. Thus, computing the former must also require time
exponential in $\Omega(m/\log^2 m)$. Since $G'$ has $\Oh(km)$ edges, this
gives a lower bound for \peval{\zrel}{w} exponential in $\Omega \left(
m/(k\log^2 m) \right)$, which is $\Omega(m/\log^2 m)$ as a function of
$m$.
\end{pfof}

\section{Remarks}
For the multivariate Tutte polynomial \ZTUT{G}{q}{w}, the current lower
bound for simple graphs is $\exp(\Omega(m/\log^3 m))$ \cite{DeHuWa10}.
One might ask whether the bounce graph construction could be used to
improve this. The weight shift corresponding to a bounce inflation is in
this case given by (for $w\neq 0$ and $q\notin \{0,-2w\}$)
\begin{equation*}
\label{eq}
\frac{q}{w_S} = \prod_{i=1}^{k}
\left(
\frac{q}{\{(q/r_i)+1\}^i-1}-1
\right) \,,\quad \text{where }r_i= \left(1+\frac{q}{w}\right)^{s_i}  -1
\end{equation*}
Unlike the expression in
Lemma~\ref{lem: bounceweights} this is not a sum of powers, so the
`$k$-byte number' analogy is remote, and there is now also a dependency on
$q$ which seems to make it difficult proving something like
Lemma~\ref{lem: lever} under the same constraints.

\bibliographystyle{abbrv}
\bibliography{refs.bib}

\newpage
\appendix

\section{Supplementary proofs}

\subsection{Lemma \ref{lem: stretch} and \ref{lem: thick}}
\label{ap: pf_thickstretch}
\begin{pfof}{Lemma \ref{lem: stretch} and \ref{lem: thick}}
We repeat the arguments from the proof in \cite[Sec.4.3]{JaVeWe90}:
Let $S$ be the set of subsets $A\subseteq E\setminus \{e\}$ that
already span the whole graph $G$, i.e.
\[
S =
\{ A\subseteq E\setminus \{e\} \, : \, \kappa(A)=1 \}
\,,
\]
and let $T$ be the set of subsets that need the edge $e$ to span
the graph:
\[
T=
\{ A\subseteq E\setminus \{e\} \, : \,
\kappa(A)=2 \;\;
\hbox{and} \;\; \kappa(A\cup\{e\})=1 \}
\,.
\]
With $w'$ denoting the weight of the edge $e$ in the
original graph $G$, \eqref{eq: r-multi} gives, for both lemmas,
\begin{equation} \label{eq: before}
\ZREL{G}{\mathbf w[e\mapsto w']} =
\sum_{A\in S} \mathbf w(A)(1+w') +
        \sum_{A\in T} \mathbf w(A)w'
\,.
\end{equation}
We will compare the partial sums here to the corresponding ones
obtained when we alter the graph. When $\varphi$ is the operation
described in Lemma~\ref{lem: stretch}, we have (with $P$
the set of edges in the path)
\[
\ZREL{\varphi(G)}{\mathbf w} =
\sum_{A\in S} \mathbf w(A) \left( \mathbf w(P) +
\sum_{i=1}^k \mathbf w(P\setminus e_i) \right) +
\sum_{A\in T} \mathbf w(A) \mathbf w(P) =
\]
\[
\sum_{A\in S} \mathbf w(A) \left( \prod_{i=1}^k w_i +
\sum_{j=1}^k\prod_{i\neq j} w_i \right) +
\sum_{A\in T} \mathbf w(A)\prod_{i=1}^k w_i
\,.
\]
Comparing corresponding sums to \eqref{eq: before}, it is easy
to check that the expressions for $w'$ and $C_p$ in
Lemma~\ref{lem: stretch} indeed make
$\ZREL{\varphi(G)}{\mathbf w} =
C_P \cdot \ZREL{G}{\mathbf w[e\mapsto w']}$.

When $\varphi$ is the operation described in Lemma~\ref{lem: thick},
we have (with $B$ the set of edges in the bundle)
\[
\ZREL{\varphi(G)}{\mathbf w} =
\sum_{A\in S} \mathbf w(A)
\left(
1+\sum_{\substack{ A' \subseteq B \\ A'\neq \emptyset }}
\mathbf w(A')
\right) +
\sum_{A\in T} \mathbf w(A)
\left(
\sum_{\substack{ A' \subseteq B \\ A'\neq \emptyset }}
\mathbf w(A')
\right)
\,,
\]
and Lemma~\ref{lem: thick} follows since
\[
\sum_{\substack{ A' \subseteq B \\ A'\neq \emptyset }}
\mathbf w(A') = \prod_{i=1}^k (w_i+1) -1
\,.
\]
\end{pfof}

\subsection{Deletion/contraction and Lemma~\ref{lem: coeff}}
\label{ap: pf_coeffs}

For an edge $e=xy \in E$, let $G\setminus e $ be the graph obtained by
deleting $e$, and let $G/e$ be the (multi)graph obtained by {\em
contracting} $e$, i.e. by identifying the end vertices $x$ and $y$ before
removing $e$ . With $w_e$ denoting the weight of edge $e$, we have the
following {\em deletion/contraction reduction} for the weighted Tutte
polynomial (see \cite[Sec. 4.3]{Sok05})
\[
\ZTUT{G}{q}{\mathbf w} =
\ZTUT{G\setminus e}{q}{\mathbf w} +
w_e \cdot \ZTUT{G/e}{q}{\mathbf w}
\,.
\]
Note that if $e\in E$ is a bridge, then $G\setminus e$ has one more
component than $G$, while in any other case both $G\setminus e$ and $G/e$
have the same number of connected components as $G$. Using the above
identity and the fact that $\ZNUL{G}{q}{\mathbf w} = q^{-1}
\ZTUT{G}{q}{\mathbf w}$, this gives:
\begin{equation} \label{eq: delcontr}
\ZNUL{G}{q}{\mathbf w} =
\begin{cases}
q \cdot \ZNUL{G\setminus e}{q}{\mathbf w} +
w_e \cdot \ZNUL{G/ e}{q}{\mathbf w}
& \text{if $e$ is a bridge},\\
  \;\;\;\;\,
\ZNUL{G\setminus e}{q}{\mathbf w} +
w_e \cdot \ZNUL{G/ e}{q}{\mathbf w}
& \text{otherwise}.
\end{cases}
\end{equation}

\begin{pfof}{Lemma~\ref{lem: coeff}}
The case $q\neq0$ is treated in Lemma~2 of \cite{DeHuWa10}.
For $q=0$, we give a reduction to Lemma~1 from the same paper.
Under our assumptions it says that $w\mapsto \ZNUL{G}{0}{\mathbf w}$
cannot be computed faster than $\exp(\Omega(m))$, where $\mathbf w$
is given by (for some set $T$ of three edges):
\begin{equation}
\mathbf w(e)=\begin{cases}
-1, & \text{ if $e\in T$},\\
w, &\text{ otherwise}.
\end{cases}
\label{eq: w def}
\end{equation}
The proof of this actually uses the restriction that
$G'=(V,E\setminus T)$ is connected, so we can assume that this is
the case. Thus, no edge in $T$ is a bridge. Three applications of
\eqref{eq: delcontr}, to delete/contract these edges, gives
\[
\ZNUL{G}{0}{\mathbf w}=
\sum_{C\subseteq\{1,2,3\}}(-1)^{|C|}\ZNUL{G_C}{0}{w}
\,,
\]
for some graphs $G_C$ of constant edge-weight $w$. These $G_C$'s
may contain loops and multiple edges from the contractions.
To address this we look at the 2-stretch \inflate{G_C}{P_3} of
each $G_C$.
This will give simple graphs of constant weight functions, and
$m$ applications of \eqref{eq: stretch} from
Corollary~\ref{cor: stretchthick} gives
\[
\ZNUL{\inflate{G_C}{P_3}}{0}{w} = (2w)^m \ZNUL{G_C}{0}{w/2}
\,.
\]
If an algorithm could compute the coefficients of $w\mapsto
\ZNUL{G}{0}{w}$ faster than $\exp(\Omega(m))$ for any simple graph
$G$, it could be used to compute eight polynomials $w \mapsto
\ZNUL{\inflate{G_C}{P_3}}{0}{w}$ (one for each subset $C$).
This would give us first $w \mapsto \ZNUL{G_C}{0}{w}$ and then
$w\mapsto \ZNUL{G}{0}{\mathbf w}$, faster than $\exp(\Omega(m))$,
which is impossible according to Lemma~1 from \cite{DeHuWa10}.
\end{pfof}

\end{document}